\documentstyle[epsf,twocolumn,aps]{revtex}
\newcommand{\eq}{\begin{equation}}
\newcommand{\ee}{\end{equation}}
\newcommand{\eqa}{\begin{eqnarray}}
\newcommand{\eea}{\end{eqnarray}}
\def\sxx{\sigma_{xx}}
\def\sxxc{\sigma_c}
\def\sxy{\sigma_{xy}}

\def\dxx{D}

\newcommand{\dt}{\partial_t}
\newcommand{\jotau}{j_0^\tau}
\newcommand{\jmu}{j_\mu}
\newcommand{\bpsi}{\bar\psi}

\newcommand{\da}{\partial_\alpha}

\newcommand{\dmu}{\partial_\mu}

\newcommand{\vimp}{V}
\newcommand{\ve}{\varepsilon}
\newcommand{\ave}{\vert\varepsilon\vert}
\newcommand{\vov}{\vert\omega\vert}
\newcommand{\vnv}{\vert\omega_n\vert}

\newcommand{\pprl}{Phys. Rev. Lett. \ } 
\newcommand{\pprb}{Phys. Rev. {B}} 
\begin{document}
\twocolumn[
\hsize\textwidth\columnwidth\hsize\csname@twocolumnfalse\endcsname
\draft
\title{Bulk Tunneling at Integer Quantum Hall Transitions}
\author{Ziqiang Wang$^{a,b}$ and Shanhui Xiong$^a$}
\address{$^a$Department of Physics, Boston College, Chestnut Hill, MA 02167}
\address{$^b$Institute for Theoretical Physics, University of California,
Santa Barbara, CA 93106-4030}
\date{\today}
\maketitle
\begin{abstract}
The tunneling into the {\em bulk} of a 2D electron system (2DES) in strong
magnetic field is studied near the integer quantum Hall transitions.
We present a nonperturbative calculation of the tunneling density
of states (TDOS) for both Coulomb and short-ranged electron-electron
interactions. In the case of Coulomb interaction, the TDOS exhibits a 
2D quantum Coulomb gap behavior, $\nu(\ve)=C_Q\ave/e^4$, with $C_Q$ a 
nonuniversal coefficient of quantum mechanical origin. 
For short-ranged interactions, we find that the TDOS at low bias follows
$\nu(\ve)/\nu (0)=1+(\ave/\ve_0)^\gamma$, where $\gamma$ is a universal 
exponent determined by the scaling dimension of short-ranged interactions.
\end{abstract}
]
The integer quantum Hall transition (IQHT) refers to the continuous,
zero-temperature phase transition between two consecutive
integer quantum Hall states in a 2DES \cite{qhe,huck}. It happens when the
Fermi energy of the 2DES moves across a critical energy located
near the center of each disorder-broadened Landau level.
On both sides of the transition, while the Hall conductivity $\sxy$ is 
integer-quantized, the dissipative conductivity $\sxx$ vanishes 
at low temperatures. At the transition, however, both $\sxx$ and $\sxy$
remain finite, supporting a critical, conducting state in two dimensions.

Theoretically, the IQHT has been studied extensively in terms of
its important, {\it noninteracting analog} in which
electron-electron interactions are ignored \cite{huck}. 
The extend to which this disordered, free electron model describes 
the IQHT in real materials depends on the effects of electronic interactions.
Recently, the stability of the noninteracting theory 
has been analyzed by calculations of the scaling dimensions of the 
interactions \cite{lwinter}. It was found that short-ranged interactions are 
irrelevant in the renormalization group sense, but the
long-ranged $1/r$-Coulomb interaction is a relevant perturbation. 
In the latter case, the noninteracting theory becomes unstable, and the 
universality class ({\it e.g.} the critical exponents) of the transition 
is expected to change based on the standard theory of critical
phenomena. The experimentally measured value of
the dynamical scaling exponent, $z=1$,
is consistent with Coulomb interaction being relevant
at the transitions \cite{engel}. However, the precise role
played by Coulomb interaction and the mechanism by which it 
governs the scaling behavior have not been understood.

The tunneling density of states (TDOS) is a simple and useful probe of 
the nature and the effects of electronic interactions. 
Recent tunneling experiments in the
integer quantum Hall regime discovered, remarkably, that the TDOS vanishes 
linearly on approaching the Fermi energy \cite{tunnelingexpt}.
Since at the transition, $\sxx$ is finite and the localization length 
$\xi$ is very large, the linear suppression of the TDOS at criticality
is expected to have a different origin than the 2D classical Coulomb gap 
behavior deep in the insulating phases. 
The results of numerical calculations of the TDOS using
Hartree-Fock approximation (HFA) of Coulomb interaction also show a linear 
Coulomb gap at {\it all} Fermi energies in the lowest Landau level
\cite{hf,thf}. Since HFA does not include screening of the exchange 
interaction, the correct behavior of the TDOS remained unclear, 
especially at the transition.

In this paper, we present a nonperturbative calculation of the
TDOS {\it at} the IQHT, taking into account the dynamical
screening of Coulomb interaction by the diffusive electrons. 
We show that the TDOS is given by,
\eq
\nu(\ve)=C_Q \ave/e^4,
\label{cgap}
\ee
at low energy $\ve$ (bias). We shall refer to Eq.~(\ref{cgap}) as the 
2D quantum Coulomb gap behavior. The coefficient $C_Q$ is not a universal
number as in the 2D classical Coulomb gap expression \cite{classical},
but rather a quantity of quantum mechanical origin. In general, it depends
on magnetic field and the microscopic details of the sample such as the 
mobility. We find, at the IQHT, 
\eq
C_Q= \sqrt{1/\pi g}
\left[1+\Phi(\sqrt{g})\right]
e^{g+{1\over 4g}\ln^2\Delta},
\label{cq}
\ee
where $g=2\pi^2\sxxc$ with $\sxxc$ the critical conductivity
in units of $e^2/\hbar$ and $\Delta=(k_f a_B)^2/(2\pi^2 g k_f l)$ 
with $a_B$ the Bohr radius and $l$ the zero-field mean free path.
$\Phi$ is the error function.
For large $\ve$, $\nu(\ve)$ crosses over to the 
perturbative diagrammatic result in strong magnetic fields 
\cite{girvin,houghton}. We also study the case of 
short-ranged interactions corresponding to experimental 
situations when the presence of nearby ground planes or the tunneling 
electrodes themselves screen the long-ranged Coulomb interaction
between electrons are large distances.
In this case, we find that the zero-bias TDOS is suppressed from
the noninteracting value but remains finite. The corrections
at finite $\ve$ follow a universal power law:
$[\nu(\ve)-\nu(0)]/\nu(0)=(\ave/\ve_0)^\gamma$ with
$\gamma=x/z$, where $x\simeq0.65$ is the scaling dimension of 
short-ranged interactions and $z=2$ \cite{lwinter}.

We begin with the partition function,
\eq
Z=\int{\cal D}[\bpsi,\psi]{\cal D}[\phi]{\cal D}[\vimp]P[\vimp]e^{-S},
\label{z}
\ee
where the imaginary time action $S=S_0+S_I$,
\eqa
S_0 &= &\int_0^\beta\!\! dt d^2r \
\bpsi\left[\dt-{1\over2m}(\da-ieA_\alpha)^2+ \vimp 
\right]\psi
\nonumber \\
S_I&=&{1\over2} \int_0^\beta dt d^2r 
d^2r^\prime\phi(r)U^{-1}(r-r^\prime)\phi(r^\prime)
\label{s} \\
&+&i\int_0^\beta\!\!dt d^2r \phi(r)\bpsi(r)\psi(r).
\nonumber
\eea
Here, $\bpsi$ and $\psi$ are the electron Grassmann fields,
$\phi$ is a scalar potential coupled to electron density, and $A_\alpha$, 
$\alpha=x,y$ is the external vector potential, $\epsilon_{\alpha\beta}
\da A_\beta=B{\hat z}$. 
Integrating out $\phi$ in Eq.~(\ref{z}) gives rise to the electron-electron
interaction $U(r-r^\prime)$ in Coulomb gauge. 
Quench average over the short-ranged impurity potential $\vimp$ is
represented in Eq.~(\ref{z}) by integrating over a Gaussian distribution
$P[\vimp]$ using the standard replica method.

The TDOS can be obtained
from the impurity averaged Green's function
$G(\tau)=\langle\psi(r,\tau)\bpsi(r,0)\rangle$,
\eq
\nu(\ve)=-{1\over\pi}{\rm Im}\int d\tau e^{i\omega_n\tau} G(\tau)
\vert_{i\omega_n\to\ve+i\eta}.
\label{nuomega}
\ee
In the imaginary time path integral, $G(\tau)$ is given by
\eq
G(\tau)=Z^{-1}\int{\cal D}[\bpsi,\psi]{\cal D}[\phi]{\cal D}[\vimp]P[\vimp]
\psi(\tau)\bpsi(0)e^{-S}.
\label{g}
\ee
Consider a U(1) gauge rotation, $\psi\to\psi e^{i\theta}$, $\bpsi\to
e^{-i\theta}$ with $\theta=\theta(r,\tau)$. In the rotated frame,
$G(\tau)$ has the same form as in Eq.~(\ref{g}), but with a transformed 
action 
\eq
S_\tau=S-i\int_0^\beta\!\! d^2r dt (\jotau a_0-\jmu a_\mu),
\label{stau}
\ee
where $a_\mu=\dmu\theta$, $\mu=\tau,x,y$, is the longitudinal
U(1) gauge field coupled to the fermion 3-current $j_\mu$, and
\eq
\jotau=\delta(r)[\Theta(t)-\Theta(t-\tau)]
\label{jotau}
\ee
is the source density (current) disturbance due to the tunneling electrons.
Because of the latter, the ground state develops a nonvanishing
charge and current density, and the saddle point of the tunneling
action $S_\tau$ is shifted from that of $S$ in Eq.~(\ref{g}).
Minimizing the action, $\partial S/\partial\theta=0$,
one finds that the induced charge and current density,
$\rho=\langle j_0\rangle$ and ${\bf J}=\langle {\bf j}\rangle$,
satisfy the continuity equation,
\eq
{\partial\over\partial t}\rho+{\bf\nabla}\cdot{\bf J}=\Gamma(r,t),
\label{continuity}
\ee
where $\Gamma(r,t)=\partial_\tau j_0^\tau=
\delta(r)[\delta(t)-\delta(t-\tau)]$ corresponds
to the process of injecting an electron at $r=0$ and time $t=0$  and
removing it at $t=\tau$. 
Since the critical conductivity is finite at the IQHT,
the charge spreading is expected to be described by (anomalous) diffusion, 
we have ${\bf J}=-\dxx({\bf \nabla}
-\gamma_H {\hat z}\times{\bf \nabla})\rho$ where $D$ is the
field-dependent diffusion coefficient and $\gamma_H=\sxy/\sxx$ 
is the Hall angle. Inserting this into 
Eq.~(\ref{continuity}), we obtain the diffusion equation,
\eq
{\partial\over\partial t}\rho-\dxx{\bf\nabla}^2\rho=\Gamma(r,t).
\label{diffusion}
\ee
Notice that $\gamma_H$ does not enter because the transverse force 
does not affect the charge spreading {\it in the bulk} of the sample.
Solving Eq.~(\ref{diffusion}) leads to
\eq
\rho(q,\omega_n)={\Gamma(\tau,\omega_n)\over\vert\omega_n\vert+
\dxx(q,\omega_n) q^2},
\label{rho}
\ee
where $\Gamma(\tau,\omega_n)=1-\exp(-i\omega_n\tau)$ and $\omega_n$ is the
boson Matsubara frequency. 

Next we go back to the tunneling action in Eq.~(\ref{stau}) and 
expand the currents around the expectation values:
$j_0=\rho+\delta j_0$ and ${\bf j}={\bf J}+\delta{\bf j}$. The fluctuating
parts now satisfy the {\it homogeneous} continuity equation:
$\partial_t \delta j_0+ {\bf\nabla}\cdot\delta{\bf j}=0$.
For convenience, we choose the unitary gauge by setting
$\phi(r,t)=a_0(r,t)$. Using Eq.~(\ref{continuity}), the tunneling
action \cite{levitov} for charge spreading becomes,
\eq
S_\tau=S_u-i\int_0^\beta\!\! dt d^2r \rho a_0,
\label{stau2} 
\ee
where the unitary gauge action $S_u$ is given by
\eqa
S_u&=& S_0+i\int_0^\beta\!\! dt d^2r \delta j_\alpha a_\alpha
\label{su} \\
&+&{1\over2}\int_0^\beta dt d^2r 
d^2r^\prime a_0(r)U^{-1}a_0(r^\prime).
\nonumber
\eea
The one-particle Green's function in Eq.~(\ref{g}) becomes,
\eq
G(\tau)=Z^{-1}\int{\cal D}[\bpsi,\psi]{\cal D}[\theta]{\cal D}[\vimp]P[\vimp]
\psi(\tau)\bpsi(0)e^{-S_\tau}.
\label{gtau}
\ee
We now quench average over the impurity potential and integrating
out the fermions for a fixed gauge configuration
\cite{ybkim,long}. Eq.~(\ref{gtau}) becomes,
\eq
G(\tau)=\int{\cal D}[\theta] G_\theta(\tau)
e^{\displaystyle i\int dt d^2r \rho a_0-S_{\rm eff}[a_\mu]},
\label{gtau2}
\ee
where $G_\theta(\tau)\equiv\langle\psi(\tau)\bpsi(0)\rangle_\theta$
is given by,
\eq
G_\theta(\tau)={\int{\cal D}[\psi,\bpsi]{\cal D}[\vimp]P[\vimp]
\psi(\tau)\bpsi(0)e^{-S_u[\psi,\bpsi,a_\mu]}\over
e^{-S_{\rm eff}[a_\mu]}},
\label{ga}
\ee
with the gauge field effective action,
\eq
e^{-S_{\rm eff}[a_\mu]}=\int{\cal D}[\psi,\bpsi]{\cal D}[\vimp]P[\vimp]
e^{-S_u[\psi,\bpsi,a_\mu]}.
\label{seff}
\ee
Thus far, all evaluations have been formal and exact, and explicitly
displayed the structure of the theory in the U(1) sector.

To calculate $G(\tau)$ in Eq.~(\ref{gtau2}), one has to make systematic
approximations to obtain $S_{\rm eff}[a_\mu]$ and $G_\theta(\tau)$.
Following essentially the formalism of Finkelshtein \cite{long,fink,alex}, 
we obtain $S_{\rm eff}$ to quadratic order 
in $a$,
\eq
S_{\rm eff}=T\sum_n\int d^2q a_0(q,\omega_n) \Pi(q,\omega_n) a_0(-q,-\omega_n).
\label{seff2}
\ee
Here $\Pi$ is the polarization function,
\eq
\Pi^{-1}(q,\omega_n)={\displaystyle U(q)\over 1+  U(q){dn\over d\mu}
{\dxx q^2\over\vnv+\dxx q^2}},
\label{pi}
\ee
which coincides with the dynamical screened Coulomb interaction
\cite{girvin,houghton}.
In Eq.~(\ref{pi}), $dn/d\mu$ is the compressibility.
The bare diffusion constant in the self-consistent
Born approximation (SCBA) is $\dxx={1\over2}r_c^2 \tau_0^{-1}$,
where $r_c=(2N+1)^{1/2}l_B$ with $l_B$ the magnetic length and
$N$ the Landau-level index. The effects of anomalous diffusion
will be discussed later. Notice that in strong field,
$\dxx$ is proportional to the field dependent scattering rate 
$1/\tau_0(B)\simeq\sqrt{\omega_c/\tau_0(0)}$.

Next we turn to $G_\theta$ in Eq.~(\ref{ga}).
In the unitary gauge, the important interference effects between the phase
of the electron wave functions have been accounted for in the tunneling
action in Eq.~(\ref{gtau2}). The amplitude fluctuations
are small for the slowly varying gauge potentials 
that dominate the dynamically screened Coulomb potential in
Eq.~(\ref{pi}) \cite{alex}. This is a unique feature
of the slow diffusive dynamics of the electrons.
We therefore neglect the $\theta$-dependence and write
$G_\theta(\tau)\approx G_0(\tau)$. The functional
integral over the gauge potential in Eq.~(\ref{gtau2}) 
can thus be carried out explicitly using the effective action in 
Eq.~(\ref{seff2}), leading to
\eq
G(\tau)=G_0(\tau)e^{W(\tau)}.
\label{gtau3}
\ee
For Coulomb interaction, $U(q)=2\pi e^2/q$ in Eq.~(\ref{pi}). Making
use of Eq.~(\ref{rho}), $W(\tau)$ has the form,
\eq
W(\tau)=-{T\over2}\sum_n\int d^2q {\vert\Gamma(\tau,\omega_n)\vert^2
\over (\vnv+\dxx q^2)^2}
{2\pi e^2\over q+ {\kappa \dxx q^2\over \vnv+\dxx q^2}}.
\label{wtau0}
\ee
where $\kappa=2\pi e^2{dn/d\mu}$ is the inverse screening length at the 
transition. Notice that $W(\tau)$ has a similar structure as the 
leading correction to the TDOS in the diagrammatic perturbation theory
in both zero \cite{aal} and strong magnetic field \cite{girvin}.
The main contribution to the $q$-integral in Eq.~(\ref{wtau0}) comes
from the region $\vnv/\dxx\kappa < q < (\vnv/\dxx)^{1/2}$.
In this region, the diffusion coefficient $\dxx$ is
a constant. The anomalous diffusion \cite{cd} only appears in the opposite
limit $\dxx q^2\gg\vnv$. As a result, $W(\tau)$ becomes,
in the $T=0$ limit,
\eq
W(\tau)={1\over4\pi^2}\int {d\omega\over\omega}{1\over\sxx}\ln(\omega\tau_s)
(1-\cos\omega\tau),
\label{wtau1}
\ee
where $\tau_s=1/\dxx \kappa^2$, and the conductivity 
$\sxx=\dxx dn/d\mu$ in units of $e^2/\hbar$
following Einstein's relation. The upper limit of the
integral in Eq.~(\ref{wtau1}) is $\hbar/\tau_0$ since one can show
$\tau_s\ll\tau_0$ near Landau level centers.
The long-time behavior of $W$ is therefore given by
$W(\tau)=(-1/8\pi^2\sxx)\ln(\tau/\tau_0)\ln(\tau/(\tau_s^2/\tau_0))$
\cite{note}.
Substituting this result into Eq.~(\ref{gtau3}), we obtain,
\eq
G(\tau)=G_0(\tau)\exp\left[-{1\over8\pi^2\sxx}
\ln\left({\tau\over\tau_0}\right)\ln\left({\tau\tau_0\over\tau_s^2}\right)
\right].
\label{gtau4}
\ee
The asymptotic behavior of $G_0(\tau)$ for large $\tau$ is
$G_0(\tau)\simeq -\nu_0/\tau$, where $\nu_0$ is the corresponding
DOS \cite{long}. In the high-field limit of the SCBA, 
it is well known that $\nu_0=(1/2\pi l_B^2)(2\tau_0)/\hbar$
and $\sxx=\nu_0\dxx\simeq (2N+1)/2\pi^2$
in the center of the N-th Landau level.
After analytical continuation to real time \cite{long},
we obtain the final result for the TDOS defined in Eq.~(\ref{nuomega}),
\eq
\nu(\ve)={2\nu_0\over\pi}\int_0^\infty dt {\sin\ave t\over t}
e^{-{1\over8\pi^2\sxx}\ln(t/\tau_0)\ln(t\tau_0/\tau_s^2)}.
\label{nu2}
\ee

We now discuss the behavior of $\nu(\ve)$ in different regimes, after
making a few remarks. (i) The quantity in the exponential 
in the above equation did not follow from an expansion in $1/\sxx$,
but rather resulted from the leading contribution dominated by the 
anomalously divergent $\ln^2$-term at long-times. Next order corrections to
the latter are of the order $\{1/\sxx,1/\sigma_{xx}^2\}\ln t/\tau_0$.
(ii) While Eq.~(\ref{nu2}) correctly captures the double-log contributions,
we have assumed that the conductivity $\sxx$ does not depend on
frequency in Eq.~(\ref{wtau1}). However, $\sxx$ can
be renormalized by localization effects 
(of leading order $(1/\sxx)\ln\omega\tau_0$ in the unitary ensemble)
and interaction effects (of leading order $\ln\omega\tau_0$
in strong magnetic field \cite{girvin,houghton}). Thus $\sxx$ assumes
the frequency-independent SCBA value only if
$-\ln(\ve\tau_0)\ll\sxx$.
In this regime, and for $-\ln(\ve\tau_0)\gg\sqrt{\sxx}$, the integral
in Eq.~(\ref{nu2}) gives,
\eq
\nu(\ve)=\nu_0\exp\left[-{1\over8\pi^2\sxx}\ln(\ave\tau_0)
\ln(\ave\tau_s^2/\tau_0)\right].
\label{nuhighf}
\ee
Expanding the exponential to leading order in $1/\sxx$, Eq.~(\ref{nuhighf})
reproduces the high-field diagrammatic perturbative result of
Girvin, et. al. \cite{girvin}. In the case of $B=0$, such
nonperturbative resummation of the perturbative double-log divergences 
was pointed out by Finkelshtein \cite{fink}, and recently
reexamined using different approaches \cite{levitov,alex,kop}.
The result of Eq.~(\ref{nuhighf}) can be regarded as an extension
of the latter to high magnetic fields.

(iii) As stated in (ii), at low-bias, {\it i.e.}
for $-\ln(\ve\tau_0)\gg\sxx$, the localization and interaction
effects lead to, in general, a frequency dependent conductivity 
$\sxx(\omega)$ in Eq.~(\ref{wtau1}). However, at the IQHT,
the critical conductivity $\sxxc$ is finite and of the order
of $e^2/h$ \cite{hf,thf,huo,wjl}. 
Thus, the validity of our analysis, {\it i.e.} the structure of the double-log
divergence at long-times, can be extended into the regime of small
$\ve$, provided that $\sxx$ in Eq.~(\ref{nu2}) is replaced
by the critical conductivity $\sxxc$.
It is important to note that because of the double-log term,
the exponential factor in Eq.~(\ref{nu2}) converges very fast such
that the TDOS becomes analytic in small $\ave$. Expanding to leading
order in $\ave$,
\eq
\nu(\ve)=\nu_0\ave{2\over\pi}\int_0^\infty\! dt 
e^{\displaystyle -{1\over8\pi^2\sxxc}\ln(t/\tau_0)\ln(t\tau_0/\tau_s^2)}.
\ee
Performing this integral, and use the fact that the compressibility is
only weakly renormalized, {\it i.e.} $dn/d\mu\simeq \nu_0$, 
we obtain the results given in Eqs.~(\ref{cgap}) and (\ref{cq})
with $\Delta=\tau_s/\tau_0$.
We thus conclude that {\em for the 1/r-Coulomb interaction, 
the bulk TDOS at IQHT exhibits a linear Coulomb gap behavior of 
quantum mechanical origin, {\it i.e.} the 2D quantum Coulomb gap}.
As a consequence, the level spacing at the transition becomes
$\Delta E\sim 1/L$, where $L$ is the size of the system,
leading to a dynamical scaling exponent $z=1$.
This behavior of the TDOS is qualitatively different from
those obtained in the clean case \cite{he} and in a weak magnetic 
field \cite{aleiner}.

We emphasize that the linear {\it quantum} Coulomb gap behavior results from
the combined effects of (i) two-dimensionality, 
(ii) 1/r---Coulomb potential, and (iii) quantum diffusion, {\it i.e.} 
a finite conductivity at $T=0$. It pertains therefore to other 
metal-insulator transitions in 2D amorphous electron systems,
provided that the critical conductivity is finite \cite{kop}. An example is
the recently discovered 2D $B=0$ metal-insulator transition \cite{serge},
although the asymptotic low temperature behavior of the critical
conductivity extracted from the experimental data in this case is still 
controversial. 

Finally, we turn to short-ranged interactions. 
It has been shown interacting potentials $1/r^p$ that decay faster than
$1/r^2$ are irrelevant and scale to zero
with scaling dimensions $x=p-2$ for $2<p<2+x_{4s}$, $x_{4s}\simeq0.65$,
and $x=x_{4s}$ for $p>2+x_{4s}$ at the IQHT \cite{lwinter}.
For simplicity, we will consider 
$U(r-r^\prime)=u\delta(r-r^\prime)$, thus $U(q)=u$ in
the screened interaction in Eq.~(\ref{pi}).
Inserting the latter into Eq.~(\ref{wtau0}), the $q$ integral 
no longer leads to the log-divergent term in $\omega$ as in Eq.~(\ref{wtau1})
due to the short range nature of the interaction.
Instead, we find \cite{long},
\eq
W_{\rm sr}(\tau)=-\int_{1/\tau}^{1/\tau_0}{\alpha\over\vov}d\omega,
\label{wtausr0}
\ee
where $\alpha$ is
a nonuniversal quantity dependent on the interacting strength. It is
given by
\eq
\alpha={1\over8\pi^2\sxxc}\lambda{2+\lambda\over (1+\lambda)^2}
\left(C+\ln\sqrt{1+\lambda}\right),
\label{alpha}
\ee
with $\lambda= u {dn\over d\mu}$, $C=1/2+1/(2-3\eta/2)$, and $\eta$
the anomalous diffusion exponent \cite{cd}.
Once again in this case, the critical conductivity 
$\sxxc$ is finite at the transition such that Eqs.~(\ref{wtausr0})
and (\ref{alpha}) represent the leading contribution to $W_{\rm sr}
(\tau)$ in the
asymptotic limit. If the interaction $u$ were a marginal perturbation,
then one would obtain, as in the Luttinger liquid case,
$W_{\rm sr}=-\alpha\ln(\tau/\tau_0)$, $G_{\rm sr}(\tau)\sim\tau^{-(1+\alpha)}$,
and $\nu(\ve)\sim \nu_0\ave^{\alpha}$. However, this is not true, because
$u$ is an irrelevant perturbation and the effective interaction scales
to zero according to $u_{\rm eff}\sim u\omega^{x/z}$ with $z=2$
at the noninteracting fixed point \cite{lwinter}. As a result, $\alpha$ obeys
a scaling form $\alpha(u,\omega)={\cal A}(u\omega^{x/z})$.
The fact that
${\cal A}(u\to0,\omega)=0$ implies, together with Eq.~(\ref{alpha}),
the leading scaling behavior $\alpha\simeq A \lambda (\omega\tau_0)^{x/z}$ with
$A=C/4\pi^2\sigma_c$.
Substituting this into Eq.~(\ref{wtausr0}), one finds that
$W_{\rm sr}(\tau)=A \lambda \gamma^{-1}[(\tau_0/\tau)^{\gamma}-1]$
where $\gamma=x/z\simeq0.32$. That $W_{\rm sr}(\tau)$ converges in the
large-$\tau$ limit is a consequence of the interaction being irrelevant,
{\it i.e.} $\gamma>0$. Thus, we find for large $\tau$, 
\eq
G_{\rm sr}(\tau)= \nu(0){1\over\tau}\exp\left[{A\lambda\over\gamma}
\left(\tau\over\tau_0\right)^{-\gamma}\right],
\label{gsr}
\ee
where $\nu(0)=\nu_0\exp(-A\lambda/\gamma)<\nu_0$.
After analytic continuation, the TDOS at small $\ve$ is given by,
\eq
\nu_{\rm sr}(\ve)=\nu(0)\left[1+\left({\ave\over\ve_0}\right)^\gamma\right],
\label{nusr}
\ee
where $\ve_0=\tau_0^{-1}(A\lambda/\gamma)^{-1/\gamma}$.
This result leads to several interesting predictions: (a) For short-ranged
interactions, the TDOS is finite and nonuniversal at zero bias 
$\nu_{\rm sr}(\ve=0)=\nu(0)\ne0$. 
(b) Since $\nu(0)\ll\nu_0$ for large $\lambda$, interactions still lead
to strong density of states suppression at low bias. (c)
The plus sign in Eq.~(\ref{nusr}) indicates that the TDOS increases
with bias $\ve$ according to a universal power law
with an initial cusp singularity for our value of $\gamma$.
These predictions can, in principle, be tested experimentally by
deliberately screening out the long-ranged Coulomb interaction
using metallic gates.

The authors thank R. Ashoori, M.~P.~A. Fisher, S.~M. Girvin, I.~A. Gruzberg,
A. Kamenev, Y.~B. Kim, D.-H. Lee, 
and N. Read  for useful discussions. 
This work was supported in part by NSF Grant No. PHY94-07194 (at ITP),
DOE Grant No. DE-FG02-99ER45747, and an award from Research Corporation.


\begin{references}
\vspace{-1.0truecm}
\bibitem{qhe} For reviews see, 
{\sl The Quantum Hall Effect},
edited by R.E. Prange and S. M. Girvin (Springer-Verlag, New York, 1990);
{\sl Perspectives in the
Quantum Hall Effects}, edited by S. Das Sarma and A. Pinczuk
(John Wiley \& Sons, New York, 1997).
\bibitem{huck} B. Huckestein, Rev. Mod. Phys. {\bf67}, 357 (1995).
\bibitem{lwinter} D-H Lee and Z. Wang, \pprl {\bf76}, 4014 (1996).
\bibitem{engel} L.~W. Engel, {\it et. al.} \pprl {\bf71}, 
2368 (1993).
\bibitem{tunnelingexpt} 
H.~B. Chan, {\it et. al.} \pprl {\bf79}, 2867 (1997).
\bibitem{hf}  S.-R.~E. Yang and A.~H. MacDonald, 
\pprl {\bf 70}, 4110 (1993); 
S.-R.~E. Yang, A.~H. MacDoanld and B. Huckestein, \pprl
{\bf 74}, 3229 (1995).
\bibitem{thf} B. Huckestein and M. Backhaus, \pprl {\bf82}, 5100 (1999).
\bibitem{classical}A.~L. Efros and B.~I. Shklovskii,
J. Phys. C{\bf8}, L49 (1975).
\bibitem{girvin}S.~M. Girvin, M. Jonson, and P.~A. Lee,
\pprb{\bf26}, 1651 (1982).
\bibitem{houghton}A. Honghton, J.~R. Senna, and S.~C. Ying,
\pprb{\bf25}, 6468 (1982).
\bibitem{levitov} L.~S. Levitov and A.~V. Shytov,
Sov. Phys. JETP Lett. {\bf66}, 214 (1997).
\bibitem{ybkim} Y.~B. Kim and X.~G. Wen, \pprb{\bf50}, 8078 (1994).
\bibitem{long} S. Xiong and Z. Wang, to be published.
\bibitem{fink} A.~M. Finkelshtein, Sov. Phys. JETP {\bf57}, 97 (1983).
\bibitem{alex} A. Kamenev and A. Andreev, cond-mat/9810191.
\bibitem{aal}B.~L. Altshuler, A.~G. Aronov, and P.~A. Lee,
\pprl {\bf44}, 1288 (1980).
%
\bibitem{note} For a general bare potential $U(r)\sim 1/r^p$,
one can show \cite{long} that for $p<2$, diffusive screening dominates
and the long-time behavior of $W(\tau)$ has the double-log form.
The case of $p>2$ is similar to the short-ranged interaction case 
discussed below.
\bibitem{cd} J.~T. Chalker and G.~J. Daniell, \pprl{\bf 61}, 593 (1998).
\bibitem{kop}P. Kopietz, \pprl{\bf81}, 2120 (1998).
\bibitem{huo} Huo, R.~E. Hetzel, and R.~N. Bhatt, \pprl{\bf70}, 481 (1993)
\bibitem{wjl} Z. Wang, B. Jovanovi\'c, D-H Lee, \pprl {\bf 77}, 4426 (1996).
\bibitem{he} S. He, P.~M. Platzman, and B.~I. Halperin,
\pprl{\bf71}, 777 (1993).
\bibitem{aleiner} I.~L. Aleiner, H.~U. Baranger, and L.~I. Glazman,
\pprl{\bf74}, 3435 (1995).
\bibitem{serge} S.~V. Kravchenko, {\it et. al.} \pprb{\bf50}, 8039 (1994);
{\it ibid}, {\bf51}, 7038 (1995).
\end{references}
\end{document}